\documentstyle[12pt]{article}

\title{LORENTZ--COVARIANT ANALYSIS \\ OF A QUANTUM SOLITON}

\author{Andrei Dubikovsky\thanks{e-mail: dubikovs@sunny.bog.msu.su} \\
{\normalsize \it Department of Physics, Moscow State University, Moscow
119899, Russia}}

\date{\normalsize Talk given at the XVIII International  Workshop \\
 on High Energy Physics and Field Theory, \\
 Protvino, Russia, June 1995}

\begin{document}

\maketitle

\begin{abstract}

A set of integral relations for rotational and translational zero modes
in the vicinity of the classical soliton solution are derived from
the particle-like properties of the latter.  The validity of
these all relations is considered for a number of soliton models
in 2+1- and 3+1-dimensions.

\end{abstract}

Theories with non-trivial classical solutions, such as the skyrmion
models of baryons \cite{1}, are object of intensive
investigations.  Quantization of hedgehog-type
configurations including translational and rotational
degrees of freedom \cite{2} leads to the quantum
Hamiltonian which contains a full bilinear form in conjugated momenta
with nontrivial couplings between different collective variables
[3--5].
However, for a particle-like classical solution one
should expect additive diagonal contributions of kinetic and
centrifugal terms to the Hamiltonian, at least to the lowest
orders in the appropriate weak coupling expansion.  Actually,
it is a part of the general problem of decoupling upon
quantization of various soliton degrees of freedom, which takes
place for any type of field models with classical solutions. In
this report we'll present a consistent general analysis of this
problem, based on the
particle-like properties of the classical solution combined with
Lorentz covariance and virial relations.
In the present report, we also give an analysis
for the soliton with spin, quantized by means of translational and
rotational collective coordinates, into corresponding representation
of the Poincar\'e group.

\vskip 3 true mm

Let us consider a field theory in $d+1$ space-time dimensions
described by the Lagrangian density ${\cal L}(\varphi)$,
which possesses a classical particle-like solution
$\varphi_{c}(x)$.  It is generally accepted, that if
in the rest frame $\varphi_{c}$ is static with finite and
localized energy density, then in quantum version of the theory
such configuration describes an extended particle.
Now we'll show, that there exists a set of
nontrivial integral relations, fulfilled by $\varphi_{c}(x)$, which
provide the validity of these assumptions.

For a given static solution $\varphi_{c}(\vec x)$ the moving one
is constructed via Lorentz boost, what results in the replacement
\equation
  x^i \to {\Lambda^{-1}}^i_{\nu}x^{\nu}
\label{1}
\endequation
in the arguments of $\varphi_{c}$, where $\Lambda^{\mu}_{\nu}$ is
the corresponding Lorentz matrix.  The momentum of the moving
solution is
\equation
  P^{\mu}=\int T^{\mu 0}
   \big( \varphi_{c} ( \vec x,x^{0}) \big) \; d \vec x,
\label{2}
\endequation
where $T^{\mu\nu}(\varphi_{c})$ is the energy--momentum tensor.
Transforming the r.h.s. in (\ref{2}) to the rest frame, one gets
\equation
  P^{\mu}= \Lambda^{\mu}_{\mu'}\Lambda^{0}_{\nu'} \int
  T^{\mu'\nu'} \big( \varphi_{c} (\vec \xi) \big) \; J \; d \vec \xi,
\label{3}
\endequation
where $J={{\Lambda}^{0}_{0}}^{-1}$ is the Jacobian of transition
from $d \vec x$ to the rest frame spatial variable. On the other hand,
the l.h.s. of (\ref{2}) should be the momentum of a particle with
the mass $M$, that is
\equation
  P^{\mu}=\Lambda^{\mu}_{0}M.
\label{4}
\endequation
{}From eqs.~(\ref{3}) and (\ref{4}) for $\nu=0$ we get
$P^{0}=M,\; P^{i}=0,$ just that we should expect for a static solution.
However for $\mu=i, \; \nu=j$ we obtain
\equation
  \int { \partial {\cal L} (\varphi_c)  \over \partial \partial^{j}
    \varphi_{c}(\vec \xi)} \; \partial_{i} \varphi_{c}(\vec \xi) \;
       d \vec \xi= M  \delta_{ij}.
\label{6}
\endequation
So we get the first set of conditions
(\ref{6}), which holds for a particle-like classical configuration
$\varphi_{c}(\vec \xi)$ in the rest frame.

Now let us consider  the orbital part of the 4-rotation tensor
(without the spin term)
\equation
  L^{\mu \nu}=\int \left( x^{\nu}T^{\mu 0}
   \big( \varphi_{c} (\vec x,x^{0}) \big) -x^{\mu}T^{\nu 0}
    \big( \varphi_{c} (\vec x,x^{0}) \big) \right) d \vec x.
\label{7}
\endequation
Analogous calculations leads to
the following relations (for definiteness,
we take $d=3$)
\equation
  \int \varepsilon_{lij} \; \xi_{i} \;
\partial_{j} \varphi_{c}(\vec \xi) {\partial {\cal L} (\varphi_c) \over
\partial \partial^{k} \varphi_{c}(\vec \xi)} \; d \vec \xi =0,
\label{10}
\endequation
since $L^{0i}$ vanish in the rest frame by assumption. This is
the second set of relations on $\varphi_{c}(\vec \xi)$,
following from the Lorentz covariance and particle-likeness
of the classical solution.

So each particle-like solution should be subject of conditions
(\ref{6}) and (\ref{10}). It should be noted, that the  relation
(\ref{4}) for $\mu=0$ reproduces nothing else but the relativistic
mass-energy relation.  For the moving $\varphi^4$-kink solution
this relation has been explicitly verified in \cite{5}, and for
the moving skyrmion --- in \cite{7}) by direct calculations.
However, the eqs.~(\ref{6}) are more general and, moreover,
the eqs.~(\ref{10}) also take place.  Note also,
that these relations, being consistent with the field equations
and conservation laws, do not be the direct consequences of the
latters, and should be considered separately.

As a direct result of these relations we get the
orthogonality of  the zero--frequency eigenfunctions in the
neighborhood of the classical particle-like solution \cite{8}.
Let us discuss the theory of a nonlinear scalar field in $3$ spatial
dimensions, described by the Lagrangian density
\equation
  {\cal L} = {1 \over 2}{(\partial_{\mu} \varphi)}^{2}- U(\varphi),
\label{11}
\endequation
which possesses a static soliton solution
\equation
\varphi_{c}(x)=u(\vec x).
\label{sol}
\endequation
In the general case the non-spherical
configuration $u(\vec x)$ yields 6 zero-frequency modes ---
three translational ones
$\ \psi_{i}(\vec x)=\partial_{i}u(\vec x)\ $
and three rotational
$\ f_{i}(\vec x)=\varepsilon_{ijk}x_{j}\partial_{k}u(\vec x)\ $.
Then from eqs.~(\ref{6}) and (\ref{10}) one immediately obtains
\eqnarray
  \int d \vec \xi \; \psi_{i}(\vec \xi) \; \psi_{j}(\vec \xi)
      & = & M\delta_{ij},
\label{14}
\\
   \int d \vec \xi \; f_{i}(\vec \xi) \; \psi_{j}(\vec \xi)
      & = & 0.
\label{15}
\endeqnarray
Further, by spatial rotations one can always achieve that
\equation
  \int d \vec \xi \; f_{i}(\vec \xi) \; f_{j}(\vec \xi)=
          \Omega_{ij}=\Omega_{i}\delta_{ij},
\label{16}
\endequation
where $\Omega_{i}$ are the moments of inertia of the classical
configuration.  Obviously, the relations (\ref{14}) and (\ref{15})
remain unchanged.

So the  particle-likeness of the classical solution results
in the diagonality of the zero-frequency scalar product matrix.
This diagonality plays an essential role in the procedure of
quantization in the vicinity of a classical soliton solution
by means of collective coordinates \cite{3,5}. Following the
conventional procedure \cite{4}, let us consider the field
$\varphi(\vec x)$ in the Schr\"odinger picture in the vicinity
of the solution $u(\vec x).$ The substitution, introducing
translational and rotational collective coordinates, reads \cite{8}
\equation
  \varphi(\vec x) = u\left( R^{-1}(\vec c)(\vec x-\vec q)\right)
            +\Phi\left( R^{-1}(\vec c)(\vec x-\vec q)\right),
\label{17}
\endequation
where $\Phi$ is the meson field, $R(\vec c)$ is the rotation matrix,
$\vec q$ and $\vec c$ are the translational and rotational
collective coordinates correspondingly.

In order to keep the number of degrees of freedom we impose on
the field $\Phi(\vec y)$ 6 subsidiary conditions, which in the
theory of a weak coupling are usually taken as linear combinations
\equation
  \int d \vec y \; N^{(\alpha)}(\vec y) \; \Phi(\vec y)=0,
       \quad \alpha=1,\ldots,6.
\label{24}
\endequation
The set  $\{ N^{(\alpha)}(\vec y) \}$ should ensure the condition
of orthogonality of the meson field $\Phi(\vec y)$ to
zero-frequency modes and is chosen as \cite{paper1}
$\ N^{(\alpha)}(\vec y)=\{\psi_{i}(\vec y)/M, \; f_{i}(\vec y)/\Omega_{i}\}$.
It is this relation, that ensures the additive form  of
the collective  coordinate part of the Hamiltonian within the
weak coupling expansion in powers of the meson fields.
Considering the condition (\ref{24}) as relation, defining $\vec q$ and
$\vec c$ as functionals of $ \varphi(\vec x)$
and calculating the conjugate momentum
$\pi(\vec x)=-i { \delta \over \delta \varphi(\vec x)}$
as a composite derivative,
we can obtain finally for the Hamiltonian
the following lowest-order expression
\equation
    H  =  M +  { {\vec K}^{2} \over 2M}
   + {1 \over 2} \sum\limits_{i} {  { I_i }^{2}
   \over \Omega_{i} } .
\label{35}
\nonumber
\endequation
In eq.~(\ref{35}) $\vec K $ and $\vec I$ are the momentum and the spin of
the field, corresponding to the rotating frame
(for details see ref.~\cite{paper1}).

It is indeed such form of the Hamiltonian, that
provides to interpret  the resulting ground state as
non-relativistic particle with the mass $M$ and moments of
inertia $\Omega_{i}$.  So the correct form of the Hamiltonian
with additive kinetic and centrifugal terms, that means the
absence of correlations between translational and rotational
degrees of freedom, is ensured by the diagonality of
zero--frequency scalar product matrix (\ref{14})--(\ref{16}).
In turn, this is a direct consequence of relations (\ref{6}) and
(\ref{10}). Note also,  that this result will be actually valid
for any field model in the neighborhood of the suitable
soliton solution.

These general considerations can be easily illustrated by concrete
models. Firstly, we consider the theory of a scalar field
in 1+1-di\-men\-si\-ons, described by  the Lagrangian density (\ref{11}).
In this case we have only one relation (\ref{6})
\equation
   \int dx \; { (\varphi'(x)) }^{2} = M,
\label{36}
\endequation
where the mass $M$ is given by
\equation
    M = \int dx \; {1 \over 2} { (\varphi'(x)) }^{2} +
         		   \int dx \; U(\varphi(x)).
\label{37}
\endequation
Performing the dilatation
$\varphi(x) \; \rightarrow \varphi(\lambda x)$
and demanding for the solution at $\lambda=1$, i.~e.
${ \left( { dM(\lambda) \over d\lambda } \right) }_{\lambda=1}=0 $,
we find the well-known Hobart--Derrick virial relation
\equation
   {1 \over 2} \int dx \; { (\varphi'(x)) }^{2} = \int dx \;
                        	     U(\varphi(x)),
\label{38}
\endequation
owing to which the ``particle-likeness condition'' (\ref{36})
is fulfilled  automatically.

     In 2+1-dimensions, the solitons in $CP_{N}$-models are
interesting examples with such particle-like properties.
As it is well-known, for $N=1$ the $CP_{N}$-model is reduced to
$O(3)$-model, described by
\equation
   {\cal L} = {1 \over 2} \partial_{\mu} \varphi^{a}
                         \partial^{\mu} \varphi^{a},
    \ \ \ \varphi^{a} \varphi^{a} = 1.
\label{40}
\endequation
The standard one-particle solution of the model is given by
\equation
  \varphi^{1}
      = \phi (r) \cos n \vartheta , \; \; \; \varphi^{2} = \phi (r) \sin n
      \vartheta , \; \; \; \varphi^{3} = { (1- \phi^{2} ) }^{1/2},
\label{42}
\endequation
where $r,\vartheta$ are polar coordinates and
$\ \phi (r) = { 4r^{n} \over r^{2n}+4 } \ $,
and describes the ``baby-skyrmion'' configuration with the topological
charge $Q=n$ and the mass $M=4 \pi Q$.  Inserting
the expression (\ref{42}) into conditions (\ref{6}) and (\ref{10})
we obtain, that the conditions of particle-likeness
for the solution (\ref{42}) are satisfied.

As a more nontrivial example, we consider the $SU(2)$-Skyrme model
in 3+1-dimensions \cite{1}, including the break--symmetry pion
mass term
\equation
    {\cal L} = -{ f^{2}_{\pi} \over 4 } \; {\rm tr} \;
          L^{2}_{\mu} + { 1 \over 32 g^{2} } \; {\rm tr} \; { [ L_{\mu} L_{\nu}
          ] }^{2}  +{ m^{2}_{\pi} \over 4 } \; {\rm tr} \; ( U+U^{+}-2 ),
\label{46}
\endequation
where, as usually, $L_{\mu}=U^{-1}\partial_{\mu}U$ is
the left chiral current and $U=\sigma+i\tau^{a}\pi^{a}$ is the
quaternion field.
Supposing the conventional ``hedgehog'' {\it Ansatz}
\equation
   \sigma=\cos \phi (r), \;\;\;\;
	    \pi^{a}={r^{a} \over r}\sin \phi (r)
\label{48}
\endequation
we find that the first
particle-likeness condition for the skyrmion is
fulfilled due to virial relation.   Further, inserting the
substitution (\ref{48}) into eqs.~(\ref{10}) we find in the same
way, that the second set of relations for the skyrmion is
provided  by the symmetry properties. So we'll get upon
quantization, that the full bilinear form considered in \cite{2},
automatically simplifies up to a diagonal construction similar
to eq.~(\ref{35}), and therefore the quantized skyrmion describes
an extended non-relativistic particle.

     Finally, we consider the 't Hooft--Polyakov monopole
for the $SU(2)$-Yang--Mills--Higgs theory, described by the
Lagrangian
\equation
   {\cal L} = -{1 \over 4} { (F^{a}_{\mu \nu}) }^{2}
      +{1 \over 2} { (D_{\mu} {\phi}^{a}) }^{2}-V(\phi)
\label{53}
\endequation
with the monopole solution
\equation
   \phi^{a} = {1 \over g}{
     r^{a} \over r^{2} } H(r), \;\;\; A^{a}_{i}= {1 \over g}
    \varepsilon_{aij}{ r_{j} \over r^{2} } (1-K(r)), \;\;\; A^{a}_{0}=0.
\label{55}
\endequation
We can find that the particle-likeness condition (\ref{6})
is fulfilled due to virial relation, and the condition (\ref{10})
is fulfilled due to symmetry properties of the expression (\ref{55}),
just as in the  case of skyrmion.

So we have  proved the validity of the ``particle-likeness''
conditions (\ref{6}) and (\ref{10}) for the most important soliton
solutions. Note, that there is a close connection
between the condition (\ref{6}) and virial relations for the static
configuration by  homogeneous dilatations.  The relations (\ref{10})
are usually fulfilled on account of symmetry properties of
classical solutions.

\vskip 3 true mm

Now we give an analysis for the soliton with spin,
into corresponding representation of the Poincar\'e group.

As a first step, we consider a nonlinear scalar field in
$3$ spatial dimensions, described by the Lagrangian density
(\ref{11})
which possesses a static soliton solution
(\ref{sol}).
According to the virial theorem,
such solutions are  unstable in more then one spatial dimension,
but for our purposes it is not so important compared to
simplicity of presentation. The angular momenta $ J^{i} $ and Lorentz
boosts $ K^{i} $ are given by \eqnarray
   J^{i} & = & \int d \vec x \; \varepsilon_{ijk}\;x^{j}\;T^{k0} =\int d
       \vec x\;\varepsilon_{ijk}\;x_{j}\;\partial_{k}\varphi \;\pi.
\label{moment}
\\
      K^{i} & = & \int d \vec x \; \left( x^{0}T^{i0}-x^{i}T^{00} \right)
	    =x^{0}P^{i}-\int d \vec x \; x^{i}\; {\cal H},
\label{boost}
\endeqnarray
In these expressions
$ \;P^{i}=\int d \vec x \; T^{i0} $ are the spatial momenta,
$ \;{\cal H}=T^{00}=\pi \dot \varphi - {\cal L} (\varphi) $ is the Hamiltonian
density
and
$ \;\pi=\partial {\cal L} / \partial \dot \varphi $ is the canonical field
momentum.

Now we put the quasiclassical soliton field $u(R^{-1}(\vec x - \vec
q))$ and its canonical momentum in the leadind qusiclassical approximation
(for details see ref.~\cite{paper2})
into corresponding Noether expressions for Lorentz generators
(\ref{moment}), (\ref{boost}) and demand
for their coincidence with the corresponding one-particle representation
of the Poincar\'e group with the same mass $M$ and spin $S$. It means, that
the Lorentz generators $J^{\mu \nu}$ should take the form \cite{barut}
\eqnarray
   J^{i} & = & \varepsilon_{ijk}\;q^{j}P^{k} +S^{i} ,
\label{momentrep}
   \\
   K^{i} & = & q^{0}P^{i}-q^{i}P^{0}
		-{\varepsilon_{ijk}\;P^{j}S^{k} \over P^{0}+M}.
\label{boostrep}
\endeqnarray

   Firstly, it is a trivial task to verify, that inserting into
eq.~(\ref{moment}) the
soliton operators, one gets identically the eq.~(\ref{momentrep}), provided
by the orthogonality conditions (\ref{14})--(\ref{16}).

Applying the same procedure to the Lorentz boost operators (\ref{boost}),
we find that
the final result is the following set of
subsidiary conditions imposed on $u(\vec x)$
\eqnarray \int d \vec \xi \;
  \xi_{i} \; \psi_{j}(\vec \xi) \; \psi_{k}(\vec \xi) & = & 0 ,
\label{newpp}
\\
  \int d \vec \xi \; \xi_{i} \;f_{j}(\vec \xi) \;
     f_{k}(\vec \xi) & = & 0 ,
\label{newff}
\\
  \int d \vec \xi \;
      \xi_{i}\; \psi_{j}(\vec \xi)\;f_{k}(\vec \xi) & = & {1 \over
       2}\varepsilon_{ijl}\;\Omega_{lk} .
\label{newpf}
\endeqnarray

These relations can be understood as a criterion of
``particle-likeness'' for the classical soliton field,
describing a spinning particle. It should be noted, that whereas the
orthogonality conditions (\ref{14})--(\ref{16}) are valid for any static
classical
solution due to the general properties of lorentz-covariance and so are
automatically consistent with equations of
motion, the relations
(\ref{newpp})--(\ref{newpf}) are more strong and restrictive. Namely,
the eqs.~(\ref{15}) and (\ref{16}) are the direct consequences from
eqs.~(\ref{newpp}) and (\ref{newpf})
correspondingly. Moreover, there might exist a static solution
$u(\vec x)$, that describes a two-soliton configuration and so
cannot be consistent with the one-particle representation of the
Poincar\'e group. In this case the relations
(\ref{newpp})--(\ref{newpf}) obviously do not hold.

On account of these general considerations we show now that the typical
hedgehog configurations of nonlinear $\sigma$-models  describe
spinning particles independently of the profile of their chiral angles.
In two spatial dimensions,  we consider the
$O(3) \ \sigma$-model, described  by the Lagrangian density
(\ref{40}), with one-particle solution (\ref{42}).
This theory is hoped to reveal the fractional
spin and statistics after adding the Hopf term.
In the case of 2 spatial dimensions we have two
translational $ \psi^{a}_{i}=\partial_{i}\varphi^{a}, \;\; (i=1,\; 2) $
and only one rotational $ \;
f^{a}=\varepsilon_{ij}\;\xi_{i}\partial_{j}\varphi^{a} \; $ zero modes for
each isospin component  $\; \varphi^{a} \;$.
By straightforward substitution it is easy to verify,
that the configuration (\ref{42}) leads to fulfilment
of conditions (\ref{newpp})--(\ref{newpf}).
Thus, the baby-skyrmion solution (\ref{42}) corresponds to the
spinning particle for any choice of the chiral angle.

In 3+1-dimensional space-time, we consider the $SU(2)$-Skyrme model
described by the Lagrangian
\equation
  {\cal L} = -{1 \over 4 } \; {\rm tr} \; L^{2}_{\mu} +
             { 1 \over 32} \; {\rm tr} \; { [ L_{\mu} L_{\nu} ] }^{2}.
\label{lagrsk}
\endequation
In terms of three independent fields $ \phi_{a} $
the expression (\ref{lagrsk}) can be rewritten as \cite{Cebula}
\equation
   {\cal L} = {1 \over 2} \; \dot \phi_{a} \; M_{ab}(\vec \phi) \;
                         \dot \phi_{b} - V(\vec \phi)
\label{lagrskM}
\endequation
(the definition of $\ M_{ab}(\vec \phi)\ $ and $\ V(\vec \phi) \ $
see in \cite{Cebula, paper2}).
{}From the Lagrangian (\ref{lagrskM}) we find the Hamiltonian  density
\equation
   {\cal H} = {1 \over 2} \; \pi_{a} \; M^{-1}_{ab}(\vec \phi) \; \pi_{b}
    + V(\vec \phi),
\label{hamskM}
\endequation
where the canonical field momentum is
$\; \pi_{a}=M_{ab}(\vec \phi)\;\dot \phi_{b} $.

The treatment of the Skyrme model differs from the theories
considered below in that point, that it contains terms of the  4th order in
derivatives. As a result, all the scalar products of the model, in
particular, the orthogonality conditions (\ref{14})--(\ref{16}) in
the vicinity of the static classical solution $ \; \phi^{a}_{c}(\vec \xi)
\; $ acquire a nontrivial integration measure. It is easy to verify, that for
 the Lagrangian (\ref{lagrskM})
the weight function in the integration measure is
$ \; M_{ab}(\phi_{c}(\vec \xi))\; $.

Now we verify the conditions of particle-likeness
for the standard hedgehog configuration
\equation
    \phi_{a}={r_{a} \over r}\phi(r).
\label{phisk}
\endequation
After some algebra we obtain, that the conditions of particle-likeness
for this configuration are
satisfied in 3 spatial dimensions as well, and once more it holds
independently of the profile of the function $\; \phi(r)$. Thus,
the soliton (\ref{phisk}) of the $SU(2)$-Skyrme model might be embedded
into the
irreductible representation of the Poincar\'e group for the particle with
spin without any restrictions on the shape of the chiral angle.

To conclude let us mention, that the present analysis can be easily
extended to other soliton models including vector fields, etc.
On the other hand, the relations
(\ref{14})--(\ref{16}) and (\ref{newpp})--(\ref{newpf}),
being independent of equations of motion, can play an
essential role of additional constraints in approximate
calculations as well. For example, they can be explored as a test for
various sample functions, used in describing the shape of the
skyrmion.
Concerning the Skyrme model, our analysis is consistent with
the well-known result \cite{skyrme},
that the spin of $SU(2)$-skyrmion can be
arbitrary.

\vskip 3 true mm

I am grateful to Prof.~K.~A.~Sveshnikov in collaboration with
whom the presented results were obtained.


\begin{thebibliography}{100}

\bibitem{1}{\it I.~Zahed and G.~E.~Brown},
      Phys. Rep. {\bf 142} (1986) 1;
           {\it G.~Holzwarth and B.~Schwesinger},
      Rep. Prog. Phys. {\bf 49} (1986) 825.
\bibitem{2}{\it H.~Verschelde and H.~Verbeke},
      Nucl. Phys. {\bf A495} (1989) 523;
           {\it H.~Yamagishi and I.~Zahed},
      Phys. Rev. {\bf D43} (1991) 891.
\bibitem{3}{\it J.-L.~Gervais, A.~Jevicki and B.~Sakita},
      Phys. Rev. {\bf D12} (1975) 1038.
\bibitem{4}{\it O.~A.~Khrustalev, A.~V.~Razumov, and A.~Yu.~Taranov},
      Nucl. Phys. {\bf B172} (1980) 44.
\bibitem{5}{\it R.~Rajaraman},
      Solitons and Instantons (North-Holland, Amsterdam, 1982).
\bibitem{7}{\it D.~P.~Cebula, A.~Klein and N.~R.~Walet},
      J.  Phys.  G:  Nucl.  Part.  Phys. {\bf 18} (1992) 499.
\bibitem{8} {\it K.~A.~Sveshnikov},
      Doctoral Dissertation (Moscow State University, Moscow, 1990).
\bibitem {paper1} {\it A.~Dubikovsky and K.~Sveshnikov},
        Phys. Lett. {\bf B317} (1993) 581.
\bibitem {barut}  {\it J.~Schwinger},
	Particles, Sources and Fields (Addison-Wesley Publishing
          Company Reading, Massachusetts, 1970).
\bibitem {Cebula} {\it D.~P.~Cebula, A.~Klein and N.~R.~Walet},
        J. Phys. G: Nucl. Part. Phys. {\bf 18} (1992) 499.
\bibitem {paper2} {\it A.~Dubikovsky and K.~Sveshnikov},
        Phys. Lett. {\bf B321} (1994) 80.
\bibitem {skyrme} {\it G.~Adkins, C.~Nappi and E.~Witten},
        Nucl. Phys.  {\bf B228} (1983) 552;
    {\it G.~Adkins and C.~Nappi},
        ibid. {\bf B233} (1984) 109.


\end{thebibliography}
\end{document}